\documentclass{article}
\usepackage[noend]{algorithmic}
\usepackage[ruled]{algorithm2e}
\usepackage{amsmath} 
\usepackage{pgf}
\usepackage{tikz}
\usetikzlibrary{arrows,automata,fit}
\usepackage[latin1]{inputenc}
\usepackage{verbatim}
\usepackage{graphicx} 
\usepackage{subfigure,tikz}
\usepackage{setspace}
\usepackage{authblk}
\usepackage{url}
\begin{document}
\title{\bf A Note on ``A polynomial-time algorithm for global value numbering"}
\author{{\normalsize Saleena Nabeezath and Vineeth Paleri}}
\affil{{\normalsize Department of Computer Science and Engineering\\National Institute of Technology Calicut, India.}\\  \texttt{{\small {\{saleena,vpaleri\}@nitc.ac.in}}}}

\date{}

\maketitle
\begin{abstract}
Global Value Numbering(GVN) is a popular method for detecting redundant computations. 
A polynomial time algorithm for GVN is presented by Gulwani and Necula(2006). Here we present two limitations of this GVN algorithm due to which detection of certain kinds of redundancies can not be done using this algorithm. The first one is concerning the use of this algorithm in detecting some instances of the classical global common subexpressions, and the second is concerning its use in the detection of some redundancies that a local value numbering algorithm will detect. We suggest improvements that enable the algorithm to detect these kinds of redundancies as well.
\end{abstract}

\maketitle

\section{Introduction}
Global Value Numbering is a well-known approach for detecting redundant computations in programs, based on equivalence among expressions. A GVN algorithm is considered to be \emph{complete} (or \emph{precise}), if it can detect all \emph{Herbrand equivalences} among program expressions. Two expressions are said to be \textit{Herbrand equivalent} (or \textit{transparent equivalent} ), if they are computed by the same operator applied to equivalent operands \cite{sumit, rosen, rks}. 

Kildall's GVN algorithm \cite{kild} is \emph{complete} in detecting  all \emph{Herbrand equivalences} among program expressions. Gulwani and Necula \cite{sumit} present a polynomial time algorithm for GVN. This uses a data structure called \textit{Strong Equivalence Dag (SED)} for representing the structured partitions of Kildall \cite{kild}. We have observed two limitations of this algorithm due to which it misses detection of some of the redundancies that are detected by Kildall \cite{kild}. In the next section, we present two examples to demonstrate the limitations. The first one is an instance of the classical global common subexpressions that Kildall detects, whereas the second one is an instance of a redundancy detected by local value numbering. We suggest possible improvements that will make the algorithm more precise.

\section{GVN algorithm by Gulwani and Necula\cite{sumit} }
\subsection{Problem 1: Join algorithm}

\begin{figure}[ht]
\centering
\begin{tikzpicture}[scale=.75, transform shape, font=\sf\small,->,>=stealth',shorten >=1pt,auto,node distance=2.5cm,thin]
  \tikzstyle{every state}=[fill=white,draw=black,text=black]
 \node [circle]       (D) [distance=0.5cm, draw=black, fill=black] {};
\node[rectangle, draw=black, text width=2.25cm,minimum height=2mm]   (F)[above  of=D,yshift=2cm]     {\small {\hspace*{0.0 in}$\hspace*{0.25 in} x:=1; \newline \hspace*{0.25 in} y:=2; $}};
\node[rectangle, draw=black, text width=2.5cm,minimum height=2mm]   (A)[above left of=D,yshift=0.5cm]     {\small {\hspace*{0.0 in}$\hspace*{0.1 in}c:=x+y; \newline p_1: $}};
  
  \node[rectangle, draw=black, text width=2.5cm,minimum height=2mm]          (C) [above right of=D,yshift=0.5cm] {\small{\hspace*{0.0 in}$\hspace*{0.1 in}d:=x+y;\newline p_2:$}};
  \node[rectangle, draw=black,text width=2.5cm,minimum height=2mm]          (E) [below of=D,yshift=-1.5cm]        {\small{$p_3:\newline\hspace*{0.1 in}e:=x+y; $}};
  \path 
  (F) edge (A)
  (F) edge (C)
  (C) edge (D)
  (A) edge (D)
  (D) edge (E);
\node[rectangle, text width=1.5cm,minimum height=2mm]          (G11) [below of=A,xshift=-1.25cm, yshift=1.25cm]        {\small{\hspace*{0.0 in}$<c,+>$}};
\node[rectangle, text width=1.5cm,minimum height=2mm]          (G12) [below of=G11,xshift=-0.75cm, yshift=1cm]        {\small{\hspace*{0.0 in}$<x,1>$}};
\node[rectangle, text width=1.5cm,minimum height=2mm]          (G13) [below of=G11,xshift=0.75cm,yshift=1cm]        {\small{\hspace*{0.0 in}$<y,2>$}};
\node[rectangle, text width=1.5cm,minimum height=2mm]         
(G14) [left of=G12,xshift=1.25cm,yshift=0.0cm]        {\small{$<d,\perp>$}};
 \path 
        (G11) edge              (G12)
	(G11) edge              (G13);
	 \node[draw,style=dotted,fit=(G11)(G12)(G13)(G14)](group){\hspace*{-1.25 in}G1};
	 
\node[rectangle, text width=7.5cm,minimum height=2mm]          (E1) [below of=G14,xshift=1.5cm, yshift=1.5cm]        {\small{\hspace*{0.0 in}$E_1:\{\ [d],\ [x,\ 1],\ [y,\ 2],\newline\hspace*{0.35 in} [c,\ x+y,\ 1+y,\ x+2,\ 1+2]\ \}$}};

\node[rectangle, text width=1.5cm,minimum height=2mm]          (G11) [below of=C,xshift=2.75cm, yshift=1.25cm]        {\small{\hspace*{0.0 in}$<d,+>$}};
\node[rectangle, text width=1.5cm,minimum height=2mm]          (G12) [below of=G11,xshift=-0.75cm, yshift=1cm]        {\small{\hspace*{0.0 in}$<x,1>$}};
\node[rectangle, text width=1.5cm,minimum height=2mm]          (G13) [below of=G11,xshift=0.75cm,yshift=1cm]        {\small{\hspace*{0.0 in}$<y,2>$}};
\node[rectangle, text width=1.5cm,minimum height=2mm]         
(G14) [left of=G12,xshift=1.25cm,yshift=0.0cm]        {\small{\hspace*{0.0 in}$<c,\perp>$}};
 \path 
        (G11) edge              (G12)
	(G11) edge              (G13);
	 \node[draw,style=dotted,fit=(G11)(G12)(G13)(G14)](group){\hspace*{-1.25 in}G2};	
\node[rectangle, text width=7.5cm,minimum height=2mm]          (E2) [below of=G14,xshift=3cm, yshift=1.5cm]        {\small{\hspace*{0.0 in}$E_2:\{\ [c],\ [x,\ 1],\ [y,\ 2] \newline\hspace*{0.35 in} [d,\ x+y,\ 1+y,\ x+2,\ 1+2]\ \}$}};

\node[rectangle, text width=3cm,minimum height=2mm]         
(G3) [above of=E,xshift=3 cm,yshift=-1.3cm]        {\small{\hspace*{0.15 in}$<x,1><y,2>\newline\hspace*{0.1 in}<d,\perp><c,\perp>$}};
 \node[draw,style=dotted,fit=(G3)](group){\hspace*{-1.55 in}G3};	\node[rectangle, text width=7.5cm,minimum height=2mm]          (E3) [below of=E1,xshift=2cm, yshift=1cm]        {\small{\hspace*{0.0 in}$E_3:\{\ [c],\ [d],\ [x,\ 1],\ [y,\ 2] \newline\hspace*{0.35 in} [x+y,\ 1+y,\ x+2,\ 1+2]\ \}$}};

 \end{tikzpicture}
\caption{\small Join of SEDs: for program point $p_i$,  $G_i$ is the SED that Gulwani and Necula \cite{sumit} computes and $E_i$ is the optimizing pool that Kildall\cite{kild} computes.}
\label{join}

\end{figure}
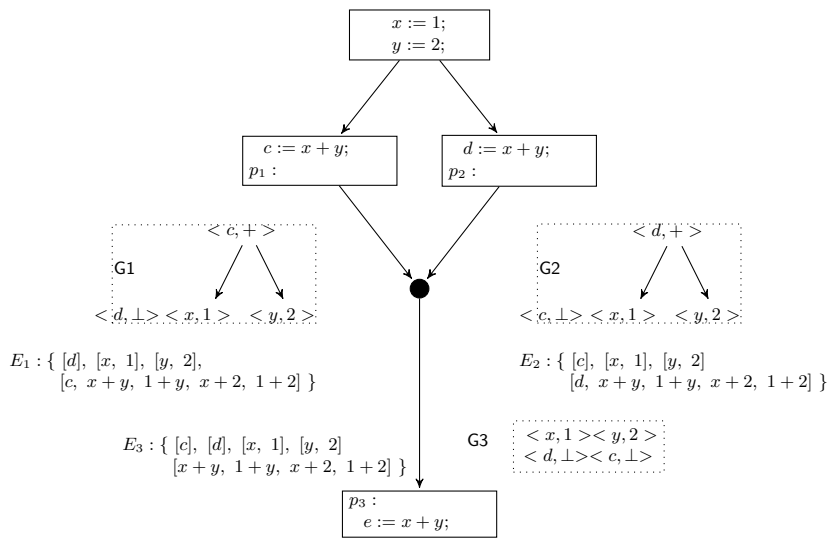
Figure \ref{join} shows four program nodes and a join point\footnote{For convenience, we use $x+y$ instead of $F(x,\ y)$}. $G_1$ and $G_2$ are the SEDs at program points $p_1$ and $p_2$ respectively. $E_1$ and $E_2$ are the structured partitions that Kildall \cite{kild} computes at these points. 
$G_3$ is the SED resulting after the join of the SEDs $G_1$ and $G_2$. The corresponding  partition in Kildall \cite{kild} is $E_3$, which is the result of the $meet$ of $E_1$ and $E_2$. 

It can be easily observed that the expression $x+y$ in the  bottommost node is redundant. Since $x+y$ is present in $E_3$, using Kildall's algorithm \cite{kild}, we can detect this redundancy. But the expression $x+y$ is not represented in the SED $G_3$, and hence the GVN algorithm by Gulwani and Necula \cite{sumit} can not detect the redundancy of $x+y$ in this example. 
\subsubsection{A solution} 
\begin{figure}[ht]
\centering
\begin{tikzpicture}[scale=.75, transform shape, font=\sf\small,->,>=stealth',shorten >=1pt,auto,node distance=2.5cm,thin]
  \tikzstyle{every state}=[fill=white,draw=black,text=black]
 \node [circle]       (D) [distance=0.5cm, draw=black, fill=black] {};
\node[rectangle, draw=black, text width=2.25cm,minimum height=2mm]   (F)[above  of=D,yshift=2cm]     {\small {\hspace*{0.0 in}$\hspace*{0.25 in} x:=1;\newline \hspace*{0.25 in} y:=2;  $}};
\node[rectangle, draw=black, text width=2.25cm,minimum height=2mm]   (A)[above left of=D,yshift=0.5cm]     {\small {\hspace*{0.0 in}$\hspace*{0.0 in}c:=x+y; \newline p_1: $}};
  
  \node[rectangle, draw=black, text width=2.25cm,minimum height=2mm]          (C) [above right of=D,yshift=0.5cm] {\small{\hspace*{0.0 in}$ \hspace*{0.0 in}d:=x+y;\newline p_2:$}};
  \node[rectangle, draw=black,text width=2.25cm,minimum height=2mm]          (E) [below of=D,yshift=-0.5cm]        {\small{\hspace*{0.0 in}$p_3:\newline\hspace*{0.1 in} e:=x+y;$}};
  \path 
  (F) edge (A)
  (F) edge (C)
  (C) edge (D)
  (A) edge (D)
  (D) edge (E);
\node[rectangle, text width=1.5cm,minimum height=2mm]          (G11) [below of=A,xshift=-1.25cm, yshift=1.25cm]        {\small{\hspace*{0.0 in}$<c,+>$}};
\node[rectangle, text width=1.5cm,minimum height=2mm]          (G12) [below of=G11,xshift=-0.75cm, yshift=1cm]        {\small{\hspace*{0.0 in}$<x,1>$}};
\node[rectangle, text width=1.5cm,minimum height=2mm]          (G13) [below of=G11,xshift=0.75cm,yshift=1cm]        {\small{\hspace*{0.0 in}$<y,2>$}};
\node[rectangle, text width=1.5cm,minimum height=2mm]         
(G14) [left of=G12,xshift=1.25cm,yshift=0.0cm]        {\small{$ <d,\perp>$}};
 \path 
        (G11) edge              (G12)
	(G11) edge              (G13);
	 \node[draw,style=dotted,fit=(G11)(G12)(G13)(G14)](group){\hspace*{-1.25 in}G1};
	 
\node[rectangle, text width=1.5cm,minimum height=2mm]          (G11) [below of=C,xshift=2.75cm, yshift=1.25cm]        {\small{\hspace*{0.0 in}$<d,+>$}};
\node[rectangle, text width=1.5cm,minimum height=2mm]          (G12) [below of=G11,xshift=-0.75cm, yshift=1cm]        {\small{\hspace*{0.0 in}$<x,1>$}};
\node[rectangle, text width=1.5cm,minimum height=2mm]          (G13) [below of=G11,xshift=0.75cm,yshift=1cm]        {\small{\hspace*{0.0 in}$<y,2>$}};
\node[rectangle, text width=1.5cm,minimum height=2mm]         
(G14) [left of=G12,xshift=1.25cm,yshift=0.0cm]        {\small{\hspace*{0.0 in}$<c,\perp>$}};
 \path 
        (G11) edge              (G12)
	(G11) edge              (G13);
	 \node[draw,style=dotted,fit=(G11)(G12)(G13)(G14)](group){\hspace*{-1.25 in}G2};	 
	 

\node[rectangle, text width=1.5cm,minimum height=2mm]          (G11) [right of=E,xshift=3.2cm, yshift=1.25cm]        {\small{\hspace*{0.0 in}$<\phi,+>$}};
\node[rectangle, text width=1.5cm,minimum height=2mm]          (G12) [below of=G11,xshift=-0.75cm, yshift=1cm]        {\small{\hspace*{0.0 in}$<x,1>$}};
\node[rectangle, text width=1.5cm,minimum height=2mm]          (G13) [below of=G11,xshift=0.75cm,yshift=1cm]        {\small{\hspace*{0.0 in}$<y,2>$}};
\node[rectangle, text width=2.5cm,minimum height=2mm]         
(G14) [left of=G12,xshift=0.5cm,yshift=0.0cm]        {\small{\hspace*{0.0 in}$<d,\perp>\ <c,\perp>$}};
 \path 
        (G11) edge              (G12)
	(G11) edge              (G13);
	 \node[draw,style=dotted,fit=(G11)(G12)(G13)(G14)](group){\hspace*{-1.25 in}G3};	
 \end{tikzpicture}
\caption{\small Join of SEDs: pairwise intersection of nodes}
\label{example1}

\end{figure}
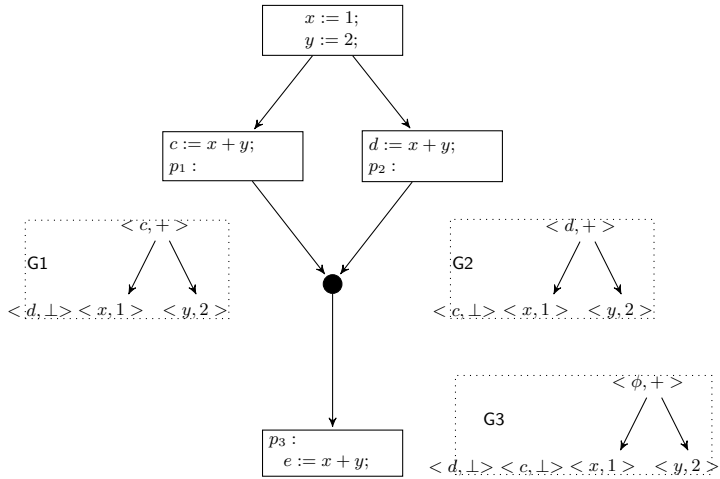

At a join point, the $meet$ operation in Kildall does intersection of every pair of classes that have at least one common $expression$, whereas the $Join$ algorithm in \cite{sumit} computes intersection of only those SED nodes having at least one common $variable$ (see line 3 of the $Join$ algorithm: \emph{for each variable $x \in T \dots$ Intersect($Node_{G_1}(x),\ Node_{G_2}(x));$}). Hence, a solution that will enable  the algorithm to detect these kinds of redundancies is to modify the $Join$ algorithm in such a way that, it computes the intersection of \emph{every pair of nodes} in the two SEDs. In Figure \ref{example1}, SED $G_3$ shows the result of computing $Join$ using the proposed method. The intersection of $<c,+>$ in $G_1$ and  $<d,+>$ in $G_2$  results in the node $<\phi,+>$ in $G_3$, which represents $x+y$ and its equivalent expressions. It may be noted that nodes like $<\phi,+>$, having empty set of variables are considered \emph{unnecessary} by Gulwani and Necula \cite{sumit}. But in fact these are \emph{necessary} (as will be shown in the next section) and hence the proposed method will retain such nodes.

\subsection{Problem 2: Removal of SED nodes}
\begin{figure}[ht]
\centering
\begin{tikzpicture}[scale=.75, transform shape, font=\sf\small,->,>=stealth',shorten >=1pt,auto,node distance=2.5cm,thin]
  \tikzstyle{every state}=[fill=white,draw=black,text=black]
\node[rectangle, draw=black, text width=2.75cm,minimum height=2mm]   (A)[]     {\small {\hspace*{0.2 in}$x:=1; y:=2;  \newline\hspace*{0.2 in} c:=x+y;\newline p_1:\newline \hspace*{0.2 in}x:=3; y:=4; \newline \hspace*{0.2 in} c :=5;\newline p_2:\newline \hspace*{0.2 in}a:=1; b:=2;\newline\hspace*{0.2 in} d:=a+b;$}};
  
\node[rectangle, text width=1.5cm,minimum height=2mm]          (G11) [right of=A,xshift=3.5cm, yshift=2cm]        {\small{\hspace*{0.0 in}$<c,+>$}};
\node[rectangle, text width=1.5cm,minimum height=2mm]          (G12) [below of=G11,xshift=-0.75cm, yshift=1cm]        {\small{\hspace*{0.0 in}$<x,1>$}};
\node[rectangle, text width=1.5cm,minimum height=2mm]          (G13) [below of=G11,xshift=0.75cm,yshift=1cm]        {\small{\hspace*{0.0 in}$<y,2>$}};
\node[rectangle, text width=3.5cm,minimum height=2mm]         
(G14) [left of=G12,xshift=1.25cm,yshift=0.0cm]        {\small{$<a,\perp>\newline <b,\perp> <d,\perp>$}};
 \path 
        (G11) edge              (G12)
	(G11) edge              (G13);
	 \node[draw,style=dotted,fit=(G11)(G12)(G13)(G14)](group){\hspace*{-1.25 in}G1};
	 
\node[rectangle, text width=5cm,minimum height=2mm]         
(G2) [right of=A,xshift=2.25cm, yshift=-0.85cm]        {\small{\hspace*{0.15 in}$\hspace*{0.25 in}<a,\perp> <b,\perp><d,\perp>\newline \hspace*{0.35 in} <x,3> <y,4> <c,5>$}};
 \node[draw,style=dotted,fit=(G2)](group){\hspace*{-1.5 in}G2};	

\node[rectangle, text width=1.5cm,minimum height=2mm]          (G11) [right of= G2,xshift=5.75cm, yshift=0.5cm]        {\small{\hspace*{0.0 in}$<\phi,+>$}};
\node[rectangle, text width=1.5cm,minimum height=2mm]          (G12) [below of=G11,xshift=-0.75cm, yshift=1cm]        {\small{\hspace*{0.0 in}$<\phi,1>$}};
\node[rectangle, text width=1.5cm,minimum height=2mm]          (G13) [below of=G11,xshift=0.75cm,yshift=1cm]        {\small{\hspace*{0.0 in}$<\phi,2>$}};
\node[rectangle, text width=4.75cm,minimum height=2mm]         
(G14) [left of=G12,xshift=1cm,yshift=0.0cm]        {\small{\hspace*{0.0 in}$<a,\perp> <b,\perp>\newline<d,\perp><x,3>\newline<y,4> <c,5>$}};
 \path 
        (G11) edge              (G12)
	(G11) edge              (G13);
	 \node[draw,style=dotted,fit=(G11)(G12)(G13)(G14)](group){\hspace*{-2.4 in}G2'};	 
	 
 \end{tikzpicture}
\caption{\small Removal of ``unnecessary" nodes: $G_1$ and $G_2$ are the SEDs at points $p_1$ and $p_2$ respectively. $G_2'$ is the required SED at $p_2$.}
\label{removal}
\end{figure}
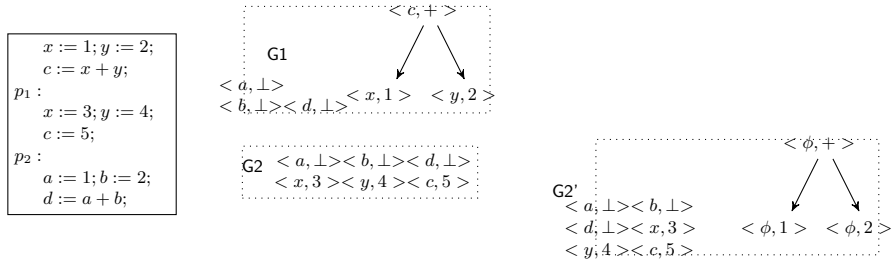

Figure \ref{removal} shows a basic block with a redundant expression $a+b$. Here the local value numbering algorithm \cite{appel} will assign the same value number to the expressions $x+y$ and $a+b$ and hence $a+b$ can be identified as redundant\footnote{according to the definition of Herbrand equivalence in \cite{rks}, $x+y$ and $a+b$ are not Herbrand equivalent}. But Gulwani and Necula \cite{sumit} can not identify this redundancy because of the following reasons: in section 3.1 of Gulwani and Necula \cite{sumit}, it is stated that \emph{the transfer functions may yield SEDs with unnecessary nodes, and these unnecessary nodes may be removed} (a node is considered unnecessary when all its ancestor nodes or all its descendant nodes have an empty set of variables). Also, it is stated in section 5.1 that \emph{the data structure (SED) represents only those partition classes explicitly that have at least one variable.} Accordingly $G_2$ is the SED computed by the algorithm at program point  $p_2$. In fact, the required SED is $G_2'$ which includes the three nodes $<\phi,1>$,  $<\phi,2>$ and $<\phi,+>$. But such nodes are considered to be unnecessary and hence will be removed by the algorithm. 
It can be observed that the node $<\phi,+>$ in $G_2'$ represents the expression $1+2$ and the same value is computed by $a+b$. With the removal of this node, it is not possible to detect that the expression $a+b$ is redundant. 
\subsubsection{The solution}
From the above example, it is clear that the problem is due to the removal of some nodes, which the algorithm considers as unnecessary. The simple solution is to retain all such nodes. In that case, for the above example,  the SED reaching the input point of $d:=a+b$ will have a node representing the expression $a+b$, indicating that this expression is redundant.
\section{Conclusion}
The GVN algorithm by Kildall was formulated with the aim of detecting \emph{common sub expressions}. An optimization using this algorithm will subsume \emph{local value numbering} also. The first example shown is an instance of the classical \emph{common sub expression elimination} and the second is an instance of \emph{local value numbering}. Hence the suggested modifications are necessary to make use of the GVN algorithm by Gulwani and Necula \cite{sumit} in compiler code optimization.


\end{document}